# Derivation and analysis of the nonlinear boundary conditions at the deformable interface between two fluids

Ivan V. Kazachkov[1,2]

[1]Dept of Information Technologies and Data Analysis,
Nizhyn Gogol State University, UKRAINE, http://www.ndu.edu.ua
[2]Dept of Energy Technology, Royal Institute of Technology, Stockholm, 10044, SWEDEN,
Ivan.Kazachkov@energy.kth.se, http://www.kth.se/itm/inst?l=en_UK

**Abstract**
   The boundary conditions at the deformable interface between two contacting fluids are derived for the general case of the large-amplitude perturbations. The interface is modeled as perturbed free boundary that evolves in time, and the non-linear description is performed and analyzed in a wide range of physical situations. The differential equations of the interfacial motion thus obtained may be useful in research of the non-linear development of the classical hydrodynamic instabilities. They should play an important role in the understanding of the hydrodynamic phenomena associated with the flows involving complex interfacial evolution including parametric control of the boundaries in continua (for example, with electromagnetic field or/and vibration).

## 1. Introduction

   The modeling of fluid interfaces (boundaries between different immiscible fluids, fluid and gas, etc.) presents tremendous challenges, which are caused by the interplay between interface dynamics and the fluid flow in contacting regions. Fluid interfacial motion induced by surface tension plays important role in diverse industrial and natural processes [1-12]. Examples of such phenomena were studied in touch with capillarity [1, 2, 7], low gravity [3], cavitation [4], hydrodynamic and hydromagnetic stability [5, 7, 9, 10], reactive flows [6], crystallization [8], interfacial transport [11, 12] , and many others.

   The current studies of fluid flows with dynamic interfaces between two (or more) layers are devoted to the following problems:
   - surface instabilities,
   - parametric control of the interfaces with electromagnetic fields, vibrations, etc.,
   - governing equations for the interfaces in diverse physical situations: boiling, evaporation, crystallization, etc.

   In this paper, a "macroscopic" description is introduced when two fluids are separated by the so-called interface, which means a surface with zero thickness. And the local dynamic equilibrium conditions of a two-fluid system (fluid 1 and fluid 2 as shown in the figure) on the moving surface of any shape and amplitude are discussed. This situation is a key point for a lot of the problems for the flows with free boundaries and interfaces [8-15]. Therefore, the boundary conditions for the dynamic interface between two fluids are formulated and analyzed.

   The paper is organized as follows. In Section 2 some preliminaries as concern to boundary conditions are stated and the problems are discussed. Section 3 is devoted to the analysis of kinematic boundary condition. Then (Section 4) the formulation of non-linear dynamical equilibrium of the interface between two fluids is considered and different cases are analyzed in touch with possible external forces. In Section 5 a derivation of a non-linear dynamical condition is given. Section 6 deals with some limit cases including linear boundary condition for small-amplitude perturbations. In Section 7 the conclusion is given as concern to the practical application of the results obtained and some planning for further developments is done.

## 2. Preliminaries

First, some basic definitions, notations, and experimental knowledge are put together and discussed in touch with the mathematical simulation of the wave film flow and other hydrodynamic problems with evolutionary boundary interfaces.

For example, plane film flow has some kind of uncertainty as concern to boundary conditions. Assume that liquid film is spreading out on the rigid surface having the upper free surface or some interface with the other fluid as it is shown in the Figure. If film thickness is big enough, then Van-der-Waals forces can be neglected. By solution of the problem on film stability in a linear approach, one does not need to state the initial conditions. And considering a parametric excitation or suppression of a perturbation of the free boundary, normally the solution is sought in the same form using the linear superposition principle. Therefore, the fully determined boundary problem requires only the statement of the boundary conditions on a rigid surface and at the interface. The classic boundary condition on the body (rigid surface) is zero velocity. Though nowadays there are many evidences that tangential velocity may be non-zero. The question has been stated by a lot of researchers starting from Stokes (1845), Lamb (1947), Zhukovskii (1948). Later on, it was shown in the review [16] that on the nonwettable surface body surface the remarkable slip of fluid is possible. Happel and Brenner [17] considered as mostly reliable hypothesis that tangential velocity of fluid at each point of rigid surface has to be counted as proportional to shear stress at the local point, with so called coefficient of a slip friction β. They assumed that β depends only on the fluid and on the body surface properties. The conditions of a slip were also analyzed in some other papers, e.g. [18,19]. The mathematical substantiation was given [19] for the partial slip condition.

## 3. Kinematic condition

On the deformable free surface $z=a+\chi(x,g,t)$ or interface of two fluids, the kinematic boundary condition is considered in the form (continuity of normal velocity across the interface):

$$z=a+\chi, \qquad w_1 = w_2 = \frac{\partial \chi}{\partial t} + u_j \frac{\partial \chi}{\partial x} + v_j \frac{\partial \chi}{\partial y}, \qquad (1)$$

where $\{u, v, w\}$ is velocity vector, $j=1,2$ ("fluid 1" and "fluid 2", respectively), $\chi(x, y, t)$ is perturbation of the interface. In case of the free surface, the second fluid is absent and indexes $j$ are omitted.

Let us scrutinize the conditions (1) transforming them into the following form:

$$z=a+\chi(x,y,t), \quad w_1 = \frac{\partial \chi}{\partial t} + u_j \frac{\partial \chi}{\partial x} + v_j \frac{\partial \chi}{\partial y}, \quad (u_1-u_2)\frac{\partial \chi}{\partial x} + (v_1-v_2)\frac{\partial \chi}{\partial y} = 0, \qquad (2)$$

where the second equation was obtained by subtraction of two equations (1). From the second equation (2) follows that tangential velocities in both directions should be the same for "fluid 1" and "fluid 2" at the interface (continuity of tangential velocity across the interface):

$$z = a + \chi, \quad u_1 = u_2, \quad v_1 = v_2. \qquad (3)$$

Otherwise, if suppose the slipping fluids at the interface, then $u_1 \neq u_2$, $v_1 \neq v_2$, where from with account of the second equation in the equation array (2) yields

$$z = a + \chi, \qquad \frac{\partial \chi}{\partial x} = -\frac{(v_1-v_2)}{(u_1-u_2)} \cdot \frac{\partial \chi}{\partial y}. \qquad (4)$$

Thus, in case (3) there are available arbitrary perturbations by $x$ and $y$ directions. But in case of a slip at the interface, the equation (4) must be satisfied as the relationship between perturbations by coordinates $x$ and $y$. The result is quite unexpected: slip of the phases at the interface requests some relation between the perturbations while, if the fluids are not slipping at the interface, the boundary perturbations may be arbitrary. What is more, in case of a slip of fluids at the interface, there are impossible plane perturbations of the interface, because if $\partial \chi / \partial y = 0$, then $\partial \chi / \partial x = 0$.

## 4. Dynamic conditions

The dynamic conditions at the interface consist of balance of the shear and normal stresses. Therefore, on the perturbed interface, in case of immiscible fluids, the capillary forces should be also taken into account. They are big enough by the remarkable perturbations. The capillary force is expressed in the form:

$$p_\sigma = \sigma K , \qquad (5)$$

where σ is the surface tension coefficient, $K$ is the average curvature of a perturbed interface between two fluids. From the differential geometry [20], one can express $K$ at an arbitrary point of the interface in the following form:

$$K = \frac{\dfrac{\partial^2 \chi}{\partial x^2} + \dfrac{\partial^2 \chi}{\partial y^2} + \dfrac{\partial^2 \chi}{\partial x^2}\left(\dfrac{\partial \chi}{\partial y}\right)^2 - 2\dfrac{\partial \chi}{\partial x}\dfrac{\partial \chi}{\partial y}\dfrac{\partial^2 \chi}{\partial x \partial y} + \dfrac{\partial^2 \chi}{\partial y^2}\left(\dfrac{\partial \chi}{\partial x}\right)^2}{\left[1 + \left(\dfrac{\partial \chi}{\partial x}\right)^2 + \left(\dfrac{\partial \chi}{\partial y}\right)^2\right]^{3/2}} \qquad (6)$$

In a linear approach, in case of small perturbations of the interface from an equilibrium state, the correlation (6) gives the well-known Landau formula [7], which was obtained as the solution of the variation problem for the minimum of a full free surface energy. It is easy to examine that expression (6) contains in its expansion by small-amplitude perturbations only the odd-order terms. This is why the linear approximation [7] was successfully applied even in case of small-amplitude non-linear perturbations of a free surface (interface). It is an exact solution with accuracy up to the second-order terms by perturbations.

Normally, in the formulation of the dynamic equilibrium condition at the interface, in a linear approach, all forces are considered by z = a. So that only the projection on the unperturbed surface is considered. But in case of large-amplitude perturbations, the problem is substantially non-linear and requires considering the curved perturbed surface at each its point. The normal and tangential vectors, at each point of the interface, may deviate substantially from the stable equilibrium state. Therefore, all forces should be projected onto the normal and tangential vectors at each point of the curved surface.

## 5. Derivation of the nonlinear dynamical conditions at the interface

The unit vectors (one normal and two tangential) at each and every point of the deformable interface can be presented in the form [20]:

$$\mathbf{n} = \frac{\left\{-\dfrac{\partial \chi}{\partial x}, -\dfrac{\partial \chi}{\partial y}, 1\right\}}{\sqrt{1 + \left(\dfrac{\partial \chi}{\partial x}\right)^2 + \left(\dfrac{\partial \chi}{\partial y}\right)^2}}, \quad \tau_x = \frac{\left\{1, 0, \dfrac{\partial \chi}{\partial x}\right\}}{\sqrt{1 + \left(\dfrac{\partial \chi}{\partial x}\right)^2}}, \quad \tau_y = \frac{\left\{0, 1, \dfrac{\partial \chi}{\partial y}\right\}}{\sqrt{1 + \left(\dfrac{\partial \chi}{\partial y}\right)^2}} . \qquad (7)$$

Using (7) one can determine the stresses on the elementary plane having the normal unit vector **n**. For this purpose, first the following expressions of the hydrodynamic stresses are represented as follows [4]:

$$p_{nn} = n_x p_{nx} + n_y p_{ny} + n_z p_{nz}, \quad p_{\tau x} = \tau_{xx} p_{nx} + \tau_{xy} p_{ny} + \tau_{xz} p_{nz}, \quad p_{\tau y} = \tau_{yx} p_{nx} + \tau_{yy} p_{ny} + \tau_{yz} p_{nz}, \qquad (8)$$

where are:

$$p_{nx} = n_x p_{xx} + n_y p_{yx} + n_z p_{zx}, \qquad \mathbf{n} = \{n_x, n_y, n_z\},$$

$$p_{ny} = n_x p_{xy} + n_y p_{yy} + n_z p_{zy}, \qquad \tau_x = \{\tau_{xx}, \tau_{xy}, \tau_{xz}\}, \qquad (9)$$

$$p_{nz} = n_x p_{xz} + n_y p_{yz} + n_z p_{zz}, \qquad \tau_y = \{\tau_{yx}, \tau_{yy}, \tau_{yz}\}.$$

Then the following well-known expressions for the stress tensor are needed [4]:

$$p_{xx} = -p + 2\mu \frac{\partial u}{\partial x}, \quad p_{yy} = -p + 2\mu \frac{\partial v}{\partial y}, \quad p_{zz} = -p + 2\mu \frac{\partial w}{\partial z}, \qquad (10)$$

$$p_{xy} = p_{yx} = \mu\left(\frac{\partial u}{\partial y} + \frac{\partial v}{\partial x}\right), \quad p_{yz} = p_{zy} = \mu\left(\frac{\partial w}{\partial y} + \frac{\partial v}{\partial z}\right), \quad p_{xz} = p_{zx} = \mu\left(\frac{\partial w}{\partial x} + \frac{\partial u}{\partial z}\right),$$

where the indexes $j=1,2$ are omitted because the expressions for $j=1$ and $j=2$ are the same. Here $p$ is the pressure, $\mu$ is the dynamic viscosity coefficient. Now substituting the equations (9), (10) into (8), with account of (7), results in the following relations for the normal and shear stresses at the perturbed non-linear interface:

$$p_{nn} = -p + \frac{2\mu}{\sqrt{1 + \left(\frac{\partial \chi}{\partial x}\right)^2 + \left(\frac{\partial \chi}{\partial y}\right)^2}} \left\{ \left[\left(\frac{\partial \chi}{\partial x}\right)^2 - 1\right]\frac{\partial u}{\partial x} + \left[\left(\frac{\partial \chi}{\partial y}\right)^2 - 1\right]\frac{\partial v}{\partial y} + \right.$$

$$\left. + \left(\frac{\partial u}{\partial y} + \frac{\partial v}{\partial x}\right)\frac{\partial \chi}{\partial x}\frac{\partial \chi}{\partial y} - \left(\frac{\partial u}{\partial z} + \frac{\partial w}{\partial x}\right)\frac{\partial \chi}{\partial x} - \left(\frac{\partial v}{\partial z} + \frac{\partial w}{\partial y}\right)\frac{\partial \chi}{\partial y} \right\}, \qquad (11)$$

$$p_{\tau x} = \frac{\mu}{\sqrt{1 + \left(\frac{\partial \chi}{\partial y}\right)^2}} \frac{1}{\sqrt{1 + \left(\frac{\partial \chi}{\partial x}\right)^2 + \left(\frac{\partial \chi}{\partial y}\right)^2}} \left\{ \left(\frac{\partial u}{\partial z} + \frac{\partial w}{\partial x}\right)\left[1 - \left(\frac{\partial \chi}{\partial x}\right)^2\right] - \right.$$

$$\left. -2\left(2\frac{\partial u}{\partial x} + \frac{\partial v}{\partial y}\right)\frac{\partial \chi}{\partial x} - \left(\frac{\partial u}{\partial y} + \frac{\partial v}{\partial x}\right)\frac{\partial \chi}{\partial y} - \left(\frac{\partial v}{\partial z} + \frac{\partial w}{\partial y}\right)\frac{\partial \chi}{\partial x}\frac{\partial \chi}{\partial y} \right\}$$

The equations for $p_{\tau x}$ and $p_{\tau y}$ are symmetric with regards to the variables $x$ and $y$. Therefore, here only $p_{\tau x}$ was written. Then the dynamic equilibrium of the interface may be expressed as the following general conditions:

$$[p_{nn}]_2^1 + [\rho]_2^1 g\chi n_z = \sigma K, \quad [\vec{p}_\tau]_2^1 + [\rho]_2^1 g\chi \tau_z = 0, \qquad (12)$$

where $p_\tau = \{p_{\tau x}, p_{\tau y}\}$, $\tau_z = \{\tau_{xz}, \tau_{yz}\}$ are the 2-D vectors. The assignment $[\ ]_2^1$ means a jump of the corresponding parameter of the fluid at the interface, e.g. $[p]_2^1 = p_1 - p_2$. Here $\rho$ is density, $g$ is acceleration due to the gravity. In equation (12) only the vertical components of the normal and tangential vectors of the tangential plate at each point of the interface are considered. This is due to the fact that the case considered only one mass force is taken into account (gravitational), which is acting in the vertical direction. The first condition in (12) expresses the equilibrium of normal stresses while the other two conditions express the equilibrium of shear stresses in two perpendicular directions in the tangential plate. In general case, the gravitation may be directed arbitrary, as well as some other volumetrically distributed forces (electromagnetic, for example) may be present. Then, instead of (12) yield the following equations:

$$[p_{nn}]_2^1 + [\rho]_2^1 gn\chi + [f]_2^1 n\chi = \sigma K,$$
$$[p_{\tau x}]_2^1 + [\rho]_2^1 g\tau_x \chi + [f]_2^1 \tau_x \chi = 0, \qquad (13)$$
$$[p_{\tau y}]_2^1 + [\rho]_2^1 g\tau_y \chi + [f]_2^1 \tau_y \chi = 0.$$

For example, if the lower fluid is electro conductive one under the electromagnetic field with vertical component $H_z(x,y,t)$, then the term $0.5\mu_m H_z^2 n_z$ [14] appears in the first equation (13) to the left. Here $\mu_m$ is the magnetic permeability. If the other fluid is electro conductive while the first one is non-conductive, the inverse situation comes into being and the sign changes. Then, if the liquids are moving on the surface of some vibrating plate, the problem might be considered in the same way, supposing that the inertial coordinate system is touched with the vibrating plate. In this case, $g + g_v$ will replace $g$, where $g_v$ is the acceleration due to vibration, e.g. by given vibration amplitude $A_g$ and frequency $\omega$, $|g_v| = A_g \cos\omega t$, where $t$ is time. And the unit vector for the acceleration due to vibration is $g_v / |g_v|$.

Substitution of (6), (9) - (11) in the boundary conditions (13) results in the boundary conditions for general case, for example, in case of electro conductive fluid located below the non-conductive one, under vertical electromagnetic field [14] yields

- normal to the interface:

$$\left( p_2 - p_1 \right)\left[ 1 + \left(\frac{\partial \chi}{\partial x}\right)^2 + \left(\frac{\partial \chi}{\partial y}\right)^2 \right]^{\frac{3}{2}} + 2\left[ 1 + \left(\frac{\partial \chi}{\partial x}\right)^2 + \left(\frac{\partial \chi}{\partial y}\right)^2 \right]\left\{ \frac{1}{2}\left[ (\rho_1 - \rho_2) g \chi + \frac{1}{2}\mu_m H_z^2 \right] + \right.$$

$$\left[ \mu_1\left( \frac{\partial u_1}{\partial x}\frac{\partial \chi}{\partial x} - \frac{\partial u_1}{\partial z} - \frac{\partial w_1}{\partial x} \right) + \mu_2\left( \frac{\partial u_2}{\partial z} + \frac{\partial w_2}{\partial x} - \frac{\partial u_2}{\partial x}\frac{\partial \chi}{\partial x} \right) \right]\frac{\partial \chi}{\partial x} + \left[ \mu_2\left( \frac{\partial v_2}{\partial z} + \frac{\partial w_2}{\partial y} \right) - \mu_1\left( \frac{\partial v_1}{\partial z} + \frac{\partial w_1}{\partial y} \right) \right]\frac{\partial \chi}{\partial y} +$$

$$\left[ \mu_1\left( \frac{\partial v_1}{\partial x} + \frac{\partial u_1}{\partial y} \right) - \mu_1\left( \frac{\partial v_2}{\partial x} + \frac{\partial u_2}{\partial y} \right) \right]\frac{\partial \chi}{\partial x}\frac{\partial \chi}{\partial y} + \left( \mu_1 \frac{\partial v_1}{\partial y} - \mu_2 \frac{\partial v_2}{\partial y} \right)\left(\frac{\partial \chi}{\partial y}\right)^2 + \mu_1 \frac{\partial w_1}{\partial z} - \mu_2 \frac{\partial w_2}{\partial z} \right\} =$$

$$\sigma\left[ \frac{\partial^2 \chi}{\partial x^2} + \frac{\partial^2 \chi}{\partial y^2} + \frac{\partial^2 \chi}{\partial x^2}\left(\frac{\partial \chi}{\partial y}\right)^2 - 2\frac{\partial \chi}{\partial x}\frac{\partial \chi}{\partial y}\frac{\partial^2 \chi}{\partial x \partial y} + \frac{\partial^2 \chi}{\partial y^2}\left(\frac{\partial \chi}{\partial x}\right)^2 \right],$$

(14)

- tangential to the interface in *x*-direction:

$$\left[ 1 - \left(\frac{\partial \chi}{\partial x}\right)^2 \right]\left[ \mu_1\left( \frac{\partial w_1}{\partial x} + \frac{\partial u_1}{\partial z} \right) - \mu_2\left( \frac{\partial w_2}{\partial x} + \frac{\partial u_2}{\partial z} \right) \right] + \left\{ 2\left[ \mu_1\left( \frac{\partial w_1}{\partial z} - \frac{\partial u_1}{\partial x} \right) + \right.\right.$$

$$\left. + \mu_2\left( \frac{\partial u_2}{\partial x} - \frac{\partial w_2}{\partial z} \right) \right] + \left[ \mu_2\left( \frac{\partial w_2}{\partial y} + \frac{\partial v_2}{\partial z} \right) - \mu_1\left( \frac{\partial w_1}{\partial y} + \frac{\partial v_1}{\partial z} \right) \right]\frac{\partial \chi}{\partial y} \left\} \frac{\partial \chi}{\partial x} + \right.$$

$$+ \left[ \mu_2\left( \frac{\partial v_2}{\partial x} + \frac{\partial u_2}{\partial y} \right) - \mu_1\left( \frac{\partial v_1}{\partial x} + \frac{\partial u_1}{\partial y} \right) \right]\frac{\partial \chi}{\partial y} + \left[ (\rho_1 - \rho_2) g \chi + \frac{1}{2}\mu_m H_z^2 \right]$$

$$\times \left\{ \left[ 1 + \left(\frac{\partial \chi}{\partial x}\right)^2 + \left(\frac{\partial \chi}{\partial y}\right)^2 \right]\left[ 1 + \left(\frac{\partial \chi}{\partial x}\right)^2 \right] \right\}^{\frac{1}{2}} \frac{\partial \chi}{\partial x} = 0,$$

(15)

- and tangential to the interface in *y*-direction:

$$\left[ 1 - \left(\frac{\partial \chi}{\partial y}\right)^2 \right]\left[ \mu_1\left( \frac{\partial v_1}{\partial z} + \frac{\partial w_1}{\partial y} \right) - \mu_2\left( \frac{\partial v_2}{\partial z} + \frac{\partial w_2}{\partial y} \right) \right] + \left\{ 2\left[ \mu_1\left( \frac{\partial w_1}{\partial z} - \frac{\partial v_1}{\partial y} \right) + \mu_2\left( \frac{\partial v_2}{\partial y} - \frac{\partial w_2}{\partial z} \right) \right] + \right.$$

$$\left[ \mu_2\left( \frac{\partial w_2}{\partial x} + \frac{\partial u_2}{\partial z} \right) - \mu_1\left( \frac{\partial w_1}{\partial x} + \frac{\partial u_1}{\partial z} \right) \right]\frac{\partial \chi}{\partial x} \left\} \frac{\partial \chi}{\partial y} + \left[ \mu_2\left( \frac{\partial u_2}{\partial y} + \frac{\partial v_2}{\partial x} \right) - \mu_1\left( \frac{\partial u_1}{\partial y} + \frac{\partial v_1}{\partial x} \right) \right]\frac{\partial \chi}{\partial x} +$$

$$\left[ (\rho_1 - \rho_2) g \chi + \frac{1}{2}\mu_m H_z^2 \right]\left\{ \left[ 1 + \left(\frac{\partial \chi}{\partial x}\right)^2 + \left(\frac{\partial \chi}{\partial y}\right)^2 \right]\left[ 1 + \left(\frac{\partial \chi}{\partial y}\right)^2 \right] \right\}^{\frac{1}{2}} \frac{\partial \chi}{\partial y} = 0.$$

(16)

The equations (14)-(16) are substantially simplified for the plane perturbations when $\partial/\partial y = 0$ or $\partial/\partial x = 0$.

Normally, study of a wave motion of a perturbed interface is done under assumption that all parameters are represented through the sum of stable parameters and perturbed values, which are supposed, in most cases, as small comparing to the corresponding unperturbed values, e.g. $v = v_0 + v'$, $p = p_0 + p'$, etc. Such a linear approximation of (14)-(16) may be got taking into account the following boundary conditions for the unperturbed system (boundary interface: $z = a = $ const):

$$z = a, \quad w_{10} = w_{20} = 0, \quad \mu_1\left(\frac{\partial u_{10}}{\partial y} + \frac{\partial v_{10}}{\partial x}\right) = \mu_2\left(\frac{\partial u_{20}}{\partial y} + \frac{\partial v_{20}}{\partial x}\right), \quad (17)$$

$$\mu_1 \frac{\partial u_{10}}{\partial z} = \mu_2 \frac{\partial u_{20}}{\partial z}, \quad \mu_1 \frac{\partial v_{10}}{\partial z} = \mu_2 \frac{\partial v_{20}}{\partial z}, \quad p_{10} = p_{20} + 2\left(\mu_1 \frac{\partial w_{10}}{\partial z} - \mu_2 \frac{\partial w_{20}}{\partial z}\right).$$

Strictly speaking, in general case, there should be $u_{10} \neq u_{20}$, $v_{10} \neq v_{20}$ (tangential slip). Because the general case is too complicated, normally different simplifications are used: linear case, non-linear approximation up to the second-order terms, etc. [12, 13]. And the correlations similar to (14) - (17) may serve as basic by derivation of simpler boundary conditions, which are based on some additional hypotheses about the physics of the processes.

## 6. Linear case

For small-amplitude perturbations, one can derive from (14) - (17) using asymptotic expansions by $x$ for all functions in the vicinity of the unperturbed surface $z = a$. It results in the following linear approximation:

$$z = a, \quad w_1 = w_2 = \frac{\partial \chi}{\partial t} + u_{j0}\frac{\partial \chi}{\partial x} + v_{j0}\frac{\partial \chi}{\partial y}; \quad (18)$$

$$p_1 = p_2 + (\rho_1 - \rho_2)g\chi + \frac{\mu_m}{2}H_z^2 + 2\left(\mu_1\frac{\partial w_1}{\partial z} - \mu_2\frac{\partial w_2}{\partial z}\right) - \sigma\left(\frac{\partial^2 \chi}{\partial x^2} + \frac{\partial^2 \chi}{\partial y^2}\right); \quad (19)$$

$$\mu_1\left(\frac{\partial w_1}{\partial x} + \frac{\partial u_1}{\partial z}\right) - \mu_2\left(\frac{\partial w_2}{\partial x} + \frac{\partial u_2}{\partial z}\right) + 2\left[2\left(\mu_2\frac{\partial u_{20}}{\partial x} - \mu_1\frac{\partial u_{10}}{\partial x}\right) + \left(\mu_2\frac{\partial v_{20}}{\partial y} - \mu_1\frac{\partial v_{10}}{\partial y}\right)\right] + \frac{\mu_m}{2}H_z^2\frac{\partial \chi}{\partial x} = 0,$$

$$\mu_1\left(\frac{\partial w_1}{\partial y} + \frac{\partial v_1}{\partial z}\right) - \mu_2\left(\frac{\partial w_2}{\partial y} + \frac{\partial v_2}{\partial z}\right) + 2\left[2\left(\mu_2\frac{\partial v_{20}}{\partial y} - \mu_1\frac{\partial v_{10}}{\partial y}\right) + \left(\mu_2\frac{\partial u_{20}}{\partial x} - \mu_1\frac{\partial u_{10}}{\partial x}\right)\right] + \frac{\mu_m}{2}H_z^2\frac{\partial \chi}{\partial y} = 0. \quad (20)$$

If subtract the equation (18) for $j = 1$ from the corresponding equation for $j = 2$, may be got

$$(u_{10} - u_{20})\frac{\partial \chi}{\partial x} + (v_{10} - v_{20})\frac{\partial \chi}{\partial y} = 0. \quad (21)$$

Equation (21) thus obtained must be satisfied for any deformation $\chi$ of the interface, and the parameters of the unperturbed system do not depend on $\chi$. Therefore $u_{10} \equiv u_{20}$, $v_{10} \equiv v_{20}$ should be. Otherwise, by $u_{10} \neq u_{20}$, $v_{10} \neq v_{20}$ yields

$$\frac{\partial \chi}{\partial x} = -\frac{v_{10} - v_{20}}{u_{10} - u_{20}}\frac{\partial \chi}{\partial y}. \quad (22)$$

As (22) shows, a slip on the interface does not tolerate arbitrary perturbations, e.g. plane in particular. In general case, from equation (1) similar correlation is as follows

$$z = a, \quad (u_1 - u)\frac{\partial \chi}{\partial x} + (v_1 - v)\frac{\partial \chi}{\partial y} = 0. \quad (23)$$

So that from kinematic condition yields that two-dimensional (plane) waves are possible if and only if the slip is absent. The correlation (23) has obvious physical explanation. If we have plane waves, e.g.

$\partial \chi / \partial x = 0$, it does not matter, whether or not there is a slip by *y*-direction and yields $\partial \chi / \partial x \neq 0$, because $u_1 = u_2$ (see Figure). But on the unperturbed interface there is possible slip of phases, and the interface in accordance with (19), may be retained smooth (without perturbations). If there is no slip ($u_{10} = u_{20}$), it may cause the perturbation of the interface $\partial \chi / \partial x \neq 0$, which is clearly understood from the physical point of view as well.

Analysis of the linear boundary conditions (18) - (20) shows that shear stress on the free surface for a liquid moving in a gas may be not small, in contradiction to what is normally assumed by many researchers. Thus, this question requires careful consideration for each specific case. The equations (18) - (20) may be simplified accounting the continuity equation at the interface:

$$\frac{\partial u_j}{\partial x} + \frac{\partial v_j}{\partial y} = -\frac{\partial w_j}{\partial z}, \quad \frac{\partial u_{jo}}{\partial x} + \frac{\partial v_{jo}}{\partial y} = -\frac{\partial w_{jo}}{\partial z}. \tag{24}$$

Then (18) - (20) transform to the following kinematic condition:

$$z = \chi, \quad w_1 = w_2 = \frac{\partial \chi}{\partial t} + u_{10}\frac{\partial \chi}{\partial x} + v_{10}\frac{\partial \chi}{\partial y}; \tag{25}$$

and dynamic conditions:

$$p_1 = p_2 + (\rho_1 - \rho_2)g\chi + \frac{\mu_m}{2}H_z^2 + 2\left[\mu_2\left(\frac{\partial u_2}{\partial x} + \frac{\partial v_2}{\partial y}\right) - \mu_1\left(\frac{\partial u_1}{\partial x} + \frac{\partial v_1}{\partial y}\right)\right] - \sigma\left(\frac{\partial^2 \chi}{\partial x^2} + \frac{\partial^2 \chi}{\partial y^2}\right),$$

$$\mu_1\left(\frac{\partial w_1}{\partial x} + \frac{\partial u_1}{\partial z}\right) - \mu_2\left(\frac{\partial w_2}{\partial x} + \frac{\partial u_2}{\partial z}\right) + 2\left[2\left(\mu_2\frac{\partial u_{20}}{\partial x} - \mu_1\frac{\partial u_{10}}{\partial x}\right) + \left(\mu_2\frac{\partial v_{20}}{\partial y} - \mu_1\frac{\partial v_{10}}{\partial y}\right)\right]\frac{\partial \chi}{\partial x} + \frac{\mu_m}{2}H_z^2\frac{\partial \chi}{\partial x} = 0, \tag{26}$$

$$\mu_1\left(\frac{\partial w_1}{\partial y} + \frac{\partial v_1}{\partial z}\right) - \mu_2\left(\frac{\partial w_2}{\partial y} + \frac{\partial v_2}{\partial z}\right) + 2\left[2\left(\mu_2\frac{\partial v_{20}}{\partial y} - \mu_1\frac{\partial v_{10}}{\partial y}\right) + \left(\mu_2\frac{\partial u_{20}}{\partial x} - \mu_1\frac{\partial u_{10}}{\partial x}\right)\right]\frac{\partial \chi}{\partial y} + \frac{\mu_m}{2}H_z^2\frac{\partial \chi}{\partial y} = 0.$$

And further, because at $z = \chi$ in (26) $w_1=w_2$, it means that all derivatives of $w_1$, $w_2$ by *x* and *y* should be equal as well. Therefore, it results in the last three equations (26) as follows

$$p_1 = p_2 + (\rho_1 - \rho_2)g\chi + \frac{\mu_m}{2}H_z^2 + 2\left[+\mu_2\left(\frac{\partial u_2}{\partial x} + \frac{\partial v_2}{\partial y}\right) - \mu_1\left(\frac{\partial u_1}{\partial x} + \frac{\partial v_1}{\partial y}\right)\right] - \sigma\left(\frac{\partial^2 \chi}{\partial x^2} + \frac{\partial^2 \chi}{\partial y^2}\right), \tag{27}$$

$$(\mu_1 - \mu_2)\frac{\partial w_1}{\partial x} + \left(\mu_1\frac{\partial u_1}{\partial z} - \mu_2\frac{\partial u_2}{\partial z}\right) + 2\left[2\left(\mu_2\frac{\partial u_{20}}{\partial x} - \mu_1\frac{\partial u_{10}}{\partial x}\right) + \left(\mu_2\frac{\partial v_{20}}{\partial y} - \mu_1\frac{\partial v_{10}}{\partial y}\right)\right]\frac{\partial \chi}{\partial x} + \frac{\mu_m}{2}H_z^2\frac{\partial \chi}{\partial x} = 0 ;$$

$$(\mu_1 - \mu_2)\frac{\partial w_1}{\partial y} + \left(\mu_1\frac{\partial v_1}{\partial z} - \mu_2\frac{\partial v_2}{\partial z}\right) + 2\left[2\left(\mu_2\frac{\partial v_{20}}{\partial y} - \mu_1\frac{\partial v_{10}}{\partial y}\right) + \left(\mu_2\frac{\partial u_{20}}{\partial x} - \mu_1\frac{\partial u_{10}}{\partial x}\right)\right]\frac{\partial \chi}{\partial y} + \frac{\mu_m}{2}H_z^2\frac{\partial \chi}{\partial y} = 0.$$

When the slip of liquids at the interface is absent, then $u_{10} = u_{20}$, $v_{10} = v_{20}$, therefore the derivatives by *x* and *y* also equate. Thus, the equations (17) yield

$$\mu_1\left(\frac{\partial u_{10}}{\partial y} + \frac{\partial v_{10}}{\partial x}\right) = \mu_2\left(\frac{\partial u_{10}}{\partial y} + \frac{\partial v_{10}}{\partial x}\right), \quad \mu_1\frac{\partial u_{10}}{\partial z} = \mu_2\frac{\partial u_{20}}{\partial z},$$

$$p_{10} = p_{20} + 2\left(\mu_1\frac{\partial w_{10}}{\partial z} - \mu_2\frac{\partial w_{20}}{\partial z}\right), \quad \mu_1\frac{\partial v_{10}}{\partial z} = \mu_2\frac{\partial v_{20}}{\partial z} . \tag{28}$$

and, with account of the second equation from (24) may be got:

$$\frac{\partial w_{10}}{\partial z} = \frac{\partial w_{20}}{\partial z}. \tag{29}$$

The first equation in (27) is satisfied only in the following two cases:

a) $\mu_1 = \mu_2$, the same viscosity or the same liquids (trivial case), $p_{10} = p_{20}$;

b) $$\frac{\partial u_{10}}{\partial y} = -\frac{\partial v_{10}}{\partial x}, \quad \frac{\partial u_{10}}{\partial x} = -\frac{\partial v_{10}}{\partial y} = -\frac{\partial w_{10}}{\partial z}. \tag{30}$$

In the plane flow (e.g. $\partial/\partial y = 0$), from the equations (30) follows:

$$\frac{\partial w_{10}}{\partial z} = -\frac{\partial u_{10}}{\partial x}, \qquad \frac{\partial w_{20}}{\partial z} = -\frac{\partial u_{10}}{\partial x},$$

$$p_{10} = p_{20} + 2(\mu_1 - \mu_2)\frac{\partial w_{10}}{\partial z} = p_{20} + 2(\mu_2 - \mu_1)\frac{\partial u_{10}}{\partial x}. \tag{31}$$

And further analysis of the equations (31) shows that by $\mu_1 = \mu_2$, the pressure in both fluids is the same so that the interface has no reason to perturb. If $\mu_1 \neq \mu_2$, the situation depends on the sign of the viscosity difference $\mu_2-\mu_1$ and on the velocity gradient $\frac{\partial u_{10}}{\partial x}\left(or\ \frac{\partial w_{10}}{\partial z}\right)$. For example, if velocity decreases by $x$ $\left(\frac{\partial u_{10}}{\partial x} < 0\right)$, and the other fluid is more viscous, the pressure in a first fluid is something lower than in a second one. Therefore, the interface has tendency to penetrate into the first fluid, and so on.

The linear boundary conditions (28) thus obtained express an equilibrium of the normal and tangential forces on the interface between two fluids. They correspond to a well-known linear conditions for such situations except the normal stresses in $x$- and $y$- directions projected on a perturbed interface. The last ones have a first order by perturbations and should be also taken into account when velocities of the fluids at the interface differ (slip). Otherwise, the above-mentioned terms are omitted, and the boundary conditions at the interface of two fluids correspond the classical ones.

## 7. Concluding remarks

In this paper, the non-linear kinematic and dynamic conditions at the interface of two fluids (free surface in limit case) were considered. The equations of a non-linear dynamic evolution of the interface have been derived and analyzed. The results may be of interest for many theoretical investigations, as well as practical applications in touch with flows having free boundaries and interfaces between two fluids, which may evolve in time under some type of a parametric action, due to instability (Kelvin-Helmholtz, Tonks-Frenkel, Rayleigh-Taylor), etc. Further work to be done is extension of the results to analysis of some problems in touch with instability and parametric control of the interfaces accounting possible real peculiarities such as, for example, phase slip at the interface and non-linearity of the deformable interface.

The derived non-linear boundary conditions were applied to show that the well-known linear boundary conditions, which may be got as a limit case from the obtained ones, coincide with the known from literature only when a slip of fluids at the interface is absent. Otherwise, if velocities of the fluids at the interface differ (slip of phases), the normal stresses in $x$- and $y$- directions projected on a perturbed interface have a first order by perturbations and should be also taken into account.

## References

[1]  Levich V.G., Physicochemical Hydrodynamics, Prentice-Hall, New York, 1962.
[2]  Lamb H., Hydrodynamics (6$^{th}$ ed.), Cambridge University Press, Cambridge, 1932.
[3]  Myshkis A.D., Babskii V.G., Kopachevskii N.D., Slobozhanin L.A., Tyuptsov A.D., Low-Gravity Fluid Mechanics, Springer Verlag, New York, 1987.

[4] Batchelor G.K., An Introduction to Fluid Dynamics, Cambridge University Press, Cambridge, 1967.

[5] Drazin P.G., Reid W.H., Hydrodynamic Stability, Cambridge University Press, Cambridge, 1981.

[6] Oran E.S., Boris J.P., Numerical Simulation of Reactive Flow, Elsevier, New York, 1987.

[7] Landau L.D., Lifshitz E.M., Fluid Mechanics, Pergamon Press, New York, 1959.

[8] Sethian J.A., Strain J.J., Comp. Phys. 98 (1992) 231.

[9] Crank J., Free and Moving Boundary Problems, Oxford University Press, Oxford, 1984.

[10] Chandrasekhar S., Hydrodynamic and Hydromagnetic Stability, Oxford University Press, Oxford, 1961.

[11] Steinchen A. (ed.), Dynamics of Multiphase Flows across Interfaces, Springer, Berlin, 1996.

[12] Shyy W., Computational Modeling for Fluid Flow and Interfacial Transport, Elsevier, Amsterdam–London–New York–Tokyo, 1994.

[13] Scriven L.E., Chem. Eng. Sci. 12 (1960) 98.

[14] Kolesnichenko A.F., Kazachkov I.V., Vodyanuk V.O., Lysak N.V., Capillary MHD flows with free surfaces, Naukova Dumka, Kiev, 1988.

[15] Schlechtendahl E.G., ZAMM 78 (1998) 841.

[16] Schnell E.J., J. Appl. Phys. 27 (1956) 1149.

[17] Happel J., Brenner H., Low Reynolds Number Hydrodynamics, Kluwer Academic Publishers, Dordrecht/Boston/London, 1983.

[18] Serrin J., In Handbuch der Physik VIII/1 (1959) 125-263.

[19] Bogoryad I.B., Dynamics of Viscous Fluid With Free Surface, Tomsk State University, Tomsk, 1980.

[20] Finn R., Equilibrium Capillary Surfaces, Springer-Verlag, New York-Berlin-Heidelberg-Tokyo, 1986.

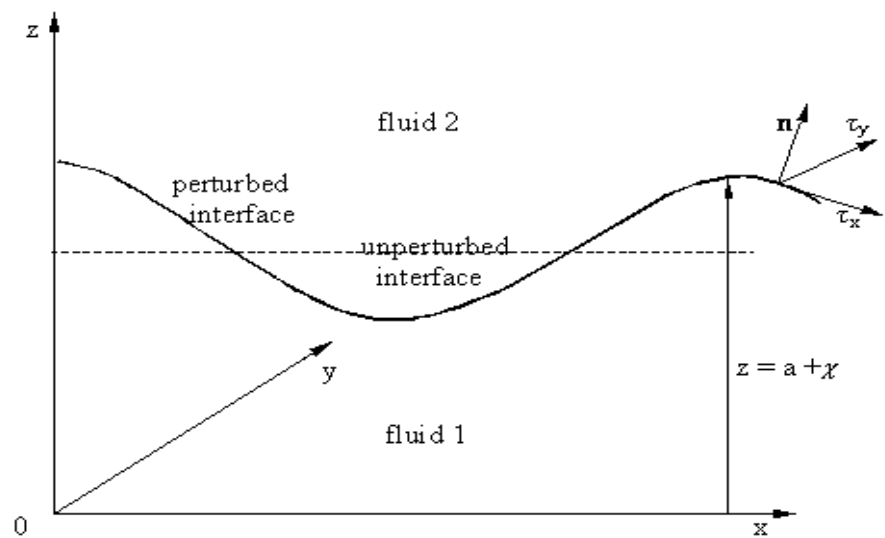

Figure. The evolution of the interface between two fluids